# Approximate Reversible Circuits for NISQ-Era Quantum Computers

Niels Gleinig
Department of Computer Science, ETH
Zurich, Switzerland
niels.gleinig@inf.ethz.ch

Tobias Rohner
Department of Computer Science, ETH
Zurich, Switzerland
tobias.rohner@math.ethz.ch

Torsten Hoefler
Department of Computer Science, ETH
Zurich, Switzerland
torsten.hoefler@inf.ethz.ch

## ABSTRACT
The synthesis approaches for quantum circuits typically aim at minimizing the number of lines or gates. Given the tight restrictions on those logical resources in physical implementations, we propose to view the problem fundamentally different: Given noisy gates and a fixed number of lines, how can we use them to perform a computation as precisely as possible? In this paper we show approximate circuits can be deployed for computations with limited resources. Performing experiments on a QC simulator, we show that under the influence of noise, approximate circuits can have lower error rates than exact circuits.

## 1 INTRODUCTION

In the near-future NISQ-era (Noisy Intermediate-Scale Quantum) of quantum computing we may have quantum computers with up to 100 qubits and strong limitations on the number of gates that we can use [24]. This has lead to much research on how to minimize the number of qubits and gates for given computational tasks [28] [12] [22] [33] [11] [17]. However, decreasing the use of one resource may exponentially increase the requirement of another resource [2]. Further, the architectures of most quantum computers impose additional restrictions on which qubits can interact with each other [38] [21]. So instead of minimizing the requirement of particular resources, in this paper we view the available resources as restricted and ask: How can we make the computation 'fit' those restrictions and handle the balance between deterministic errors and noise optimally?

Given a logic function $f$, we use an evolutionary algorithm that takes architecture restrictions as input and outputs a reversible circuit that satisfies all of the restrictions and approximates the function. The restrictions may include the number of lines (corresponding to number of qubits), number of gates (which will typically be subject to restrictions derived from the system's capacity to handle noise) and other more qualitative restrictions (for example 1-D- or 2-D-architecture specific requirements such as "only neighboring lines are allowed to interact with each other") [21]. Notice that when certain parameters such as the number of lines or gates are fixed, it may not be possible to synthesize a given function exactly. In particular, the problem: **Given $f$, $N$, $d$, and $\epsilon$, find a circuit $C$ with $N$ lines and $d$ gates that differs from $f$ on a proportion of at most $\epsilon$ inputs**, has not always a solution. Therefore our aim is not to find circuits with arbitrarily small error-rates, but circuits with error-rates that are close to the minimum that can be achieved for given restrictions.

Using this method to generate approximate reversible circuits, we proceed to investigate how the *overall quality* of the circuits changes as we increase the number of gates. Using more gates, we are able to create more complex circuits and approximations with lower deterministic error-rates. But using more gates also results in more noise and hence more non-deterministic errors. Therefore, we measure both the *deterministic error-rates* and, more importantly, the *(overall) error-rates* (the total proportion of errors observed when deploying these circuits on noisy quantum computers).

### 1.1 Reversible Circuits

Reversible gates are gates that perform a 1-to-1 mapping between input- and output-patterns. Reversible circuits are circuits that are built out of reversible gates. They are also 1-to-1 mappings and have the same number of input- and output-bits. Graphically they can be represented as a set of horizontal lines (corresponding to the input bits) and dots on the horizontal lines that are connected by vertical lines (representing gates). Given a sequence of reversible gates $c_1, \ldots, c_d$ we will denote by $[c_1, \ldots, c_d]$ the reversible circuit given by a $c_1$ gate, followed by $c_2, \ldots$ and ending on $c_d$.

In order to compute a general (classical) logic function $f$ reversibly, we need to synthesize a reversible circuit that "embeds" the function, leading to the following problem.

**Reversible Circuit Synthesis Problem:** Given a function $f : \{0, 1\}^n \rightarrow \{0, 1\}$, find a reversible circuit $C = [c_1, \ldots, c_d]$ with $N$ lines such that for any $x \in \{0, 1\}^n$ the input $(x, \vec{0}) \in \{0, 1\}^N$ is mapped by $C$ to an output that is equal to $f(x)$ on the last bit. That is, we want the output to have the form $(garbage(x), f(x))$.

Notice that $f$ and hence the value $n$ are given and define the instance of the problem. The values $N$ and $d$ are not given, but one typically tries to find a circuit for which they are as small as possible. Further, it is common to optimize for other cost metrics. In this paper we consider two different cost metrics. For the first one, we sum all gates after compilation to the IBMQ-16 Melbourne and weigh all 2-qubit gates with a factor of 10 to account for the higher error-rates (due to noise) in 2-qubit gates compared to 1-qubit gates [9]. We call this metric the **circuit cost** (cc). The second cost metric is the standard **quantum cost** (qc), defined as the number of primitive 1- and 2-qubit gates needed to represent the circuit [20].

## 1.2 Approximate Circuits

The idea of using approximate circuits in error-resilient applications has attracted much research in recent years [5][26][32]. There is a wide variety of ways to make sense of the expression "approximate circuit", arising from a multitude of possible error metrics. For functions with one output bit, we define the approximation factors in terms of the proportion of input strings that are not mapped to the correct output. For functions with "unbalanced" outputs (that is, 0 and 1 occur with very different frequencies, as for example with the function $f$ that maps $x$ to 1 iff $x$ is the binary representation of a prime number), the constant functions 0 and 1 may give the correct output on a high proportion of input strings, but fail on all inputs that make this function non-trivial. To avoid such trivial solutions that are evidently useless for computational purposes, we are interested in solutions that have both low false-positive and false-negative rates. These considerations lead to the following formal definitions of our approximation metrics. Given two functions $f : \{0,1\}^n \to \{0,1\}$ and $\hat{f} : \{0,1\}^n \to \{0,1\}$, we define the error-rate of $\hat{f}$ as an approximation of $f$ by $err(f, \hat{f}) := \frac{1}{2^n}|\{x \in \{0,1\}^n : f(x) \neq \hat{f}(x)\}|$. We define the false-negative and false-positive error-rates by

$$FN(\hat{f}, f) := \frac{|\{x \in \{0,1\}^n : \hat{f}(x)=0, f(x)=1\}|}{|\{x \in \{0,1\}^n : f(x)=1\}|},$$
$$FP(\hat{f}, f) := \frac{|\{x \in \{0,1\}^n : \hat{f}(x)=1, f(x)=0\}|}{|\{x \in \{0,1\}^n : f(x)=0\}|}.$$

We extend these definitions to general reversible circuits $\hat{C}$ by identifying $\hat{C}$ with the function $f_{\hat{C}}$ that it computes and set $err(f, \hat{C}) := err(f, f_{\hat{C}})$. In our experiments, we also consider the $F_1$-Score [8].

## 1.3 Previous Work

*1.3.1 Reversible circuit synthesis.* The exact synthesis problem for reversible circuits has been widely studied [28] [17] [12] [22] [33]. Furthermore, there exist heuristics for reduction of the resource use (mainly lines, gates, and quantum cost) of already synthesized circuits [35] [18]. It has been shown that finding the minimum number of lines needed to embed a function is coNP-hard [29], but there exists a statistical method for estimating this number [11].

*1.3.2 Evolutionary algorithms for (reversible) circuit synthesis.* EAs have been extensively used for the synthesis and optimization of logic circuits, including quantum circuits [6] [36] [13]. For an overview of EA-based synthesis of quantum circuits we recommend the surveys by Sasamal et al. [25] and Gepp and Stocks [10]. For classical circuits, EAs have also been used to find approximate circuits [26] [32] [30] [31]. However, to the best of our knowledge, in the case of quantum circuits, EAs have only been used to synthesize and optimize exact circuits, but not to reduce resource use by allowing the circuit itself to be approximate.

*1.3.3 Quantum computing.* In QC there are some approximation results like the Solovay-Kitaev theorem [7], but those are very different from the previously mentioned work: They demonstrate that there are small universal gate libraries that can approximate every unitary matrix to arbitrary precision. So they are concerned with the possibility of representing transformations that are given by arbitrary unitaries with a small family of gates (for example CNOT gates and all one qubit gates), but not with making given circuits smaller. These results are most relevant for transformations on a small number of qubits, because the number of gates needed to approximate a transformation, grows exponentially with the number of qubits. In fact, it has been shown that most unitaries cannot be *approximated* efficiently (see Knill [16] or section 4.5 of Nielsen and Chuang [23]). However, those inapproximability results do not apply to our problem, because we use a different notion of approximation. We do not approximate one particular unitary, but any of the unitaries that embed the given Boolean function, and we do not minimize the distance in terms of a Hilbert space norm, but the number of inputs that are mapped to an incorrect output. Also the work of Selinger [27], Kliuchnikov et al. [15], and Bocharov et al. [3] uses a different notion of approximation than we do, and covers only transformations on $O(1)$ lines. Khatri et al. [14] propose quantum assisted quantum compilation. This method can produce approximate circuits for a given unitary $U$. The difference to our work is that QAQC is quantum based and it needs to be given a working implementation of $U$ already.

None of the mentioned papers investigates the trade-off between deterministic error-rates and noise-induced errors.

## 1.4 Problem Formulation

In the standard synthesis problem for reversible circuits, the resources $d$ and $N$ are not given. The existing synthesis approaches typically aim at minimizing the use of at least one of them, but they do not aim at matching both to a provided amount of resources. Hence, if the function to be synthesized requires more resources than the technology can provide, the circuits produced by those approaches cannot be used at all. Further, even if the circuit complexity of the function would actually allow the function to be synthesized with the available resources, it might happen that by minimizing the use of one resource the requirement of the other resource grows beyond the amount that is available. The previously mentioned exponential time/space-trade-off in reversible computing suggests that the most useful circuits will neither use the absolute minimal number of lines nor the minimal number of gates. Therefore we study in this paper the synthesis problem with a given and fixed number of lines and gates. However, since for a given amount of those resources we cannot guarantee that there actually exists an exact circuit, we only require the circuit to be approximate.

**Approximate Reversible Circuit Synthesis Problem:** Given a function $f : \{0,1\}^n \rightarrow \{0,1\}$ and integers $N \geq n$, and $d$, find a reversible circuit on $N$ lines with $d$ gates $C = [c_1, \ldots, c_d]$ such that for any $x \in \{0,1\}^n$ the input $(x, \vec{0}) \in \{0,1\}^N$ is mapped by $C$ to an output that is equal to $\hat{f}(x)$ on the last bit, where $\hat{f}$ is a function that approximates $f$.

Evidently we want the approximation to produce as few erroneous outputs as possible and ideally, if the provided resources theoretically allow the existence of an exact circuit, we would like the circuit to be exact too. Further, we also want the computation to handle noise optimally. Increasing the number of gates $d$ we can synthesize circuits with lower deterministic error-rates but more noise. This leads to the following problem.

**Optimal Circuits under the Accuracy-Noise-Trade-Off:** Find a (possibly approximate) circuit $C$, such that executing $C$ on a noisy quantum computer gives the same output as $f$ as often as possible.

## 1.5 Results

We show experimentally that many functions can be approximated by reversible circuits that are much smaller than any known exact circuits that embed them. We find these small approximate reversible circuits using EAs. Experiments show that we can reduce both qc and cc (as defined above) by a factor of more than 16 compared to the best known exact circuit (with the same number of lines), if we allow an error-rate of 0.15. We can make this error-rate tend to zero by providing more resources. Furthermore, providing the minimal amount of resources for which exact reversible circuits are known to exist, we often find exact reversible circuits too (that is, circuits that give the correct output on all possible inputs). For some benchmark functions, the exact reversible circuits that we found had lower quantum costs than any previously known implementations. Our experiments on the IBMQ Melbourne-16 QC simulator confirm that some approximate circuits produce less errors than exact but expensive circuits. For some functions, approximate circuits produce less than half as many errors as exact circuits.

## 2 THE EVOLUTIONARY ALGORITHM

### 2.1 The Representation of Reversible Circuits

We represent a reversible circuit $C = [c_1, \cdots, c_d]$ by an array of structs representing the individual gates. These structs are composed of an enum giving the type of the gate and an array of size 3 giving the indices of the qubits taken as arguments to the instruction. For gates that have less than 3 arguments, the redundant indices are simply ignored. We make use of the 6 different gate types identity, not, controlled not, swap, Toffoli, and Fredkin.

### 2.2 The Algorithm

The evolutionary algorithm depends on the following parameters: $d$ (number of gates), $l$ (total number of lines: input lines + added lines), $S$ (number of survivors), $F$ (number of offspring per survivor), $G$ (number of generations), $b$ (batch size), $ds$ (proportion of difficult samples). The number $ds$ is a real number from $[0, 1]$. All other parameters are positive integers. In the beginning, we sample $P = S \cdot F$ reversible circuits at random. A randomly sampled circuit is built from $d$ randomly sampled gates. To generate a random gate, we first choose a gate contained in our gate library uniformly at random. Then the arguments to the chosen gate are also sampled uniformly at random from the set of all admissible argument combinations.

Now we perform the following three steps $G$ times.

- **STEP 1 (Test circuits):** Estimate the fitness of the members of the population as described in the following section.
- **STEP 2 (Select survivors):** Rank the circuits according to their estimated fitness and select the best $S$ circuits to survive and be the parents of the circuits of the next generation (truncation selection).
- **STEP 3 (Create new population):** Include each of the survivors unchanged in the new population (elite selection). Furthermore, let each survivor contribute $F - 1$ offspring to the new population, using method 2.

After having performed these steps $G$ times we arrive at a final population and choose the best circuit from it.

## 2.3 Estimation and Mutation

In the algorithm we will have to estimate the quality of circuits. We do this by testing the circuits on a set of inputs that is partly sampled uniformly at random from all possible inputs and partly from a multiset **FAILS** consisting of the inputs on which circuits from the previous generation failed. Testing the circuits on inputs from **FAILS** serves two purposes: (1) It increases genetic diversity, since circuits that are very different from the rest of the population may perform well on inputs on which many other circuits fail and hence will get an advantage by being tested on **FAILS**. (2) In the case of a very unbalanced function it prevents the population to evolve to trivial constant circuits, because those few inputs on which the function has a different value would become strongly represented in **FAILS**, giving an advantage to non-constant circuits. This testing method is summarized in Method 1.

---

**METHOD 1:** Input: A circuit $C = [c_1, \ldots, c_d]$ and a parameter $ds \in [0, 1]$. Output: An estimator $\hat{e}$ of the quality of $C$.

---

Sample $\lceil ds \cdot b \rceil$ inputs $x \in \{0, 1\}^n$ uniformly at random;
Sample $\lceil (1 - ds) \cdot b \rceil$ inputs $x \in$ **FAILS** uniformly at random;
Evaluate $C$ and $f$ on all sampled inputs;
Set $\hat{e}$ to the number of inputs on which $C$ and $f$ differ;
Include the inputs on which $C$ failed into the **FAILS** set for the next generation;
**return** $\hat{e}$;

*Note: If there are not sufficiently many elements in **FAILS**, we choose the remaining samples at random (in order to have b samples in total).*

---

Next, we describe a method to produce offspring from a given circuit. To do this we choose one of its gates at random and replace it with a new random gate (mutation). If we want the synthesized circuit to satisfy some restrictions in the end, we just need to impose those restriction on the initial population and the mutations that we do. So for example, if we can only target the CNOT gates on the first three lines, it suffices that in the initial population the circuits satisfy this restriction and all gates that we introduce later by mutations also satisfy it. This is described more precisely in method 2.

## 3 EXPERIMENTAL RESULTS

The source code we used to optimize the circuits and the plotting scripts are available on GitHub. The optimization itself is done using noiseless gates and is written in C++. To

---

**METHOD 2:** Input: A circuit $C = [c_1, \ldots, c_d]$ that abides certain architecture restrictions $R_A$. Output: Another circuit $C' = [c'_1, \ldots, c'_d]$, called a *child* of $C$.

---

Set $C' = C$;
Choose uniformly at random an integer $p \in \{1, 2, \ldots, d\}$;
Replace the gate type of $c'_p$ with one chosen uniformly at random;
**repeat**
    Replace the argument indices with three different
    integers chosen uniformly at random from $\{1, 2, \ldots, l\}$;
    //Change the control and target lines
**until** $C'$ satisfies $R_A$
**return** $C'$

---

evaluate the circuits on a noisy NISQ quantum computer, we used the Python library Qiskit initially developed at IBM. The noiseless error-rates found in this section are always exact, as we tested the circuits for every possible input. For the noisy simulations, there is a small uncertainty introduced due to Qiskit approximating the expectation value of the output by averaging over 1024 independent simulations.

### 3.1 Benchmark Functions

We examined the approximability of all 1-bit-output functions from Dmitri Maslov's benchmark page [19]. We chose this set of benchmark functions instead of the RevLib functions [34] because it contains fewer 1-bit-output functions. This allowed us to explain the experiments for each one of them in detail and yet give a representative picture by testing all of them rather than some subset. The functions are 6sym, 9sym, 2of5, 4mod5, 5mod5, and xor5. Additionally, we investigated the approximability of functions with multiple output bits using the functions NthPrime3 and NthPrime4 having 3 inputs and 5 outputs and 4 inputs and 6 outputs respectively. The output of the functions is the n-th prime indexed starting from zero.

For these benchmark functions, our EA-based computation times were high compared to exact synthesis computations. A single threaded optimization of the function 5mod5 with $l = 6, d = 20, G = 2000, S = 60, F = 100, b = 32$ took approximately 40s on an Intel Core i5-6200U CPU. Yet, the algorithm scales well as the computational cost grows at most linearly in each of the parameters.

**6sym** This function can be synthesized with 7 lines. The smallest and least expensive 7-line circuit we found in the related literature had 41 (expensive, multi-controlled) gates, qc=206 and cc= 6488, and can be seen on Maslov's benchmark page [19]. To study the approximability of this function we searched for approximate circuits with up to 30 gates. Using $l = 7, S = 60, F = 100, G = 100 \cdot d, b = 64, ds = 0.5$, we found an exact circuit using only 13 gates with qc=51, being a large improvement over the previously best known circuit. The cheapest non-constant approximate circuit we found has an error-rate 0.19, FN 0.14, FP 0.36 ($F_1 = 0.88$), and qc=45.

**9sym** This function can be synthesized with 10 lines, but all the existing 10-line implementations we found, were extremely expensive: There are two 10-line implementations on REVLIB with qc=4368 and qc=6941 and three other 10-line implementations on Maslov's benchmark page with qc=1975, qc=31819, and qc=61928. The cc of the circuit with the lowest qc was 69787. Using the parameters $d = 50, l = 10, S = 60, F = 100, G = 300, b = 180, ds = 0.5$ we found an approximate circuit with error-rate 0.15, FN 0.13, FP 0.22 ($F_1 = 0.91$), $qc = 117$, and $cc = 4171$ (this is for both $qc$ and $cc$ an improvement of more than 16x over any exact 10-line circuit).

**xor5** For this function our algorithm produced exact circuits with 5 lines and 4 gates.

**2of5** Using $l = 6$ (which is minimal), $d = 13, b = 32, G = 100 \cdot d, F = 100, S = 100, ds = 0.5$ we also obtained for this function an exact circuit. This circuit had qc= 37 which is worse than the 6-line circuits on Maslov's benchmark page. However, we also found an approximate circuit with qc= 27 and error-rate 0.13, FN 0.3 and FP 0.045 ($F_1 = 0.82$). We also found a cheaper but slightly less accurate circuit with qc= 10 and error-rate 0.19, FN 0.4 and FP 0.091 ($F_1 = 0.73$).

**4mod5** This function proved to be very easy to optimize. All non-constant circuits we found were exact with the cheapest one having a quantum cost of 9. The circuit was found using parameters $l = 6, d = 5, G = 500, S = 100, F = 100, b = 32$.

**5mod5** Using the parameters $F = 100, S = 60, G = 300, B = 180, ds = 0.5, d = 14, l = 6$, we found also for this function exact circuits with a minimal number of lines. One of the exact circuits is depicted in Figure 1a. This circuit has qc= 36, which is less than half of the qc of any 6-line implementation that can be found on Dimitri Maslov's benchmark page. It has $cc = 1628$ compared to $cc = 2174$ for the circuit with lowest qc among the 6-line circuits from Maslov's page. Furthermore, we found an approximate circuit with qc=15, error-rate 0.031, FN 0, and FP 0.04 ($F_1 = 0.92$). This makes this function one of the easiest ones to approximate. Using the parameters $B = 30, S = 1000, F = 4, G = 4, ds = 0.5$ we found a 2-gate approximation (Figure 1b) for this function which has an error-rate of $0.22, FN = 0.43$ and $FP = 0.16$ ($F_1 = 0.46$).

### 3.2 Experiments on QC Simulator

We hypothesized that using small approximate circuits we might be able to produce less noise and hence get correct results more often than with exact but expensive circuits. To test this, we ran the optimization algorithm 16 times for a

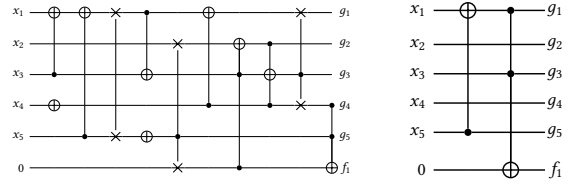

(a) Reversible circuit for 5mod5. This circuit has the minimal possible number of lines, $qc = 36$, and $cc = 1628$.

(b) 2-gate approximation for 5mod5 with $qc = 6$, $cc = 389$, error-rate $0.22, FN = 0.43$, and $FP = 0.16$.

multitude of different gate counts and evaluated the resulting circuits in a noiseless environment as well as on a simulator for the NISQ quantum computer IBMQ-16 Melbourne. The circuit evaluation was done using the Python Library Qiskit and its accurate noise models for the current quantum computing devices developed at IBM [1]. Since noisy gates are not necessarily reversible, we would have to perform a density matrix calculation in order to obtain the exact output state of our circuits. As this would be too expensive, we performed 1024 statevector simulations for each possible input of the circuits and averaged their results to get an estimate of the true expectation value of the output. In order for the circuits to be simulated on IBM's quantum computers, they first have to be transpiled to be compatible with the specific topology and basis gates of said quantum device. This introduces additional variation of the error-rates of the noisy circuits due to the transpiler being able to optimize some circuits better than others.

To visualize the data, we plotted the achieved error-rates against an estimation of the quantum cost of the circuits. The estimates were obtained by first removing the unused gates from the circuits and adding up the costs of the remaining gates. Note that this procedure provides an upper bound to the real cost of the circuits, as sometimes groups of non-primitive gates can be simplified further when working with their representation in primitive gates. Because we do not optimize circuits constrained to a given quantum cost but rather to a given amount of gates, there is most probably a different number of optimized circuits for each quantum cost. For this reason, we compute the median, minimal and maximal error-rates of all circuits having a specific cost. The median is then plotted as a line with a shaded region around it visualizing the span from the minimal to the maximal error-rates found.

The resulting errors for optimized circuits of different quantum costs for the 5mod5 function can be found in Figure 2. One can see that even for the cheapest approximate circuits, the error-rate rises to nearly $\frac{1}{2}$. However, they are still more accurate than the known exact circuits from Maslov's benchmark page. To be able to observe our hypothesis more clearly, we performed simulations where the error probabilities of the quantum gates were reduced by a factor of 10. This is a sensible choice to approximate noise models of near-term quantum computers, as IBM claims to double the quantum volume of their devices each year [9]. The resulting error-rates of 5mod5, 2of5, and NthPrime3 can be found in Figure 3 and Figure 4. We can clearly observe that the minimal errors are obtained with approximate circuits and are considerably smaller than the errors of the exact circuits from Maslov's benchmark page. Finally, we run experiments to test how resilient to noise quantum computers would need to be, so that exact circuits would perform better than approximate circuits. By scaling the error probability of the noise model we used, we are able to estimate in what range the approximate circuits are more reliable than the known exact circuits. In order to do this, we took the circuit approximation with the lowest quantum cost that was not a constant circuit and plotted its error-rate against the strength of our noise model. We did the same for the exact circuit with the lowest quantum cost found in Maslov's benchmark page. In Figure 5, these error-rates are plotted for the 2of5 and 5mod5 functions. The noise model we used is of the IBMQ-16 Melbourne quantum computer where the error probabilities are reduced by the factor plotted on the $x$-axis. We can see that for the function 2of5 where the exact circuit has a qc of approximately 30, the approximate circuit is more reliable until the noise of today's quantum computers has been reduced by 60x. In the case of the 5mod5 function, the noise of today's quantum computers would need to be reduced by over 100x in order to make the exact circuit more accurate than the approximate one.

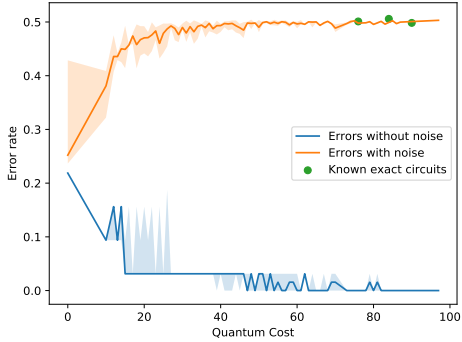

**Figure 2: Error-rates of our optimized circuits for 5mod5.**

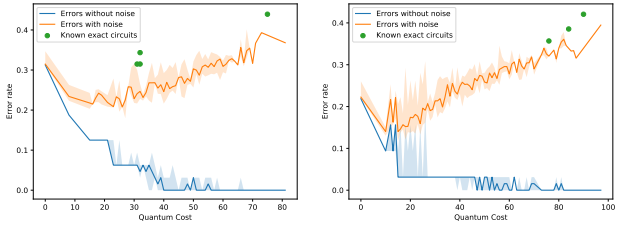

**Figure 3: Error-rates of 2of5 (left) and 5mod5 (right) plotted against qc. Notice that for 5mod5, some approximate circuits produce less than half as many errors as the exact circuits.**

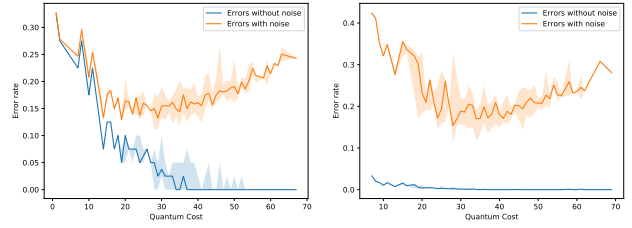

**Figure 4: Error-rates of NthPrime3 function plotted against qc using two different error metrics: Uniform weight of errors in output bits (left) and a weight of $2^i$ for errors in the $i$-th output bit (right).**

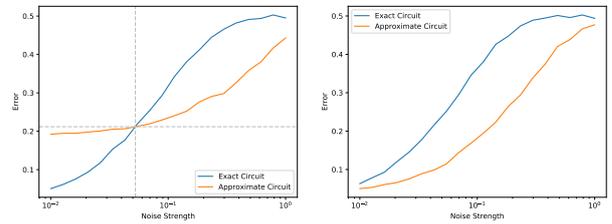

**Figure 5: Error-rates of the 2of5 (left) and 5mod5 (right) plotted against the strength of the noise simulated.**

## 4 DISCUSSION

To understand how and for which types of algorithms the methods presented in this paper could be useful, we need to discuss some of the properties of the synthesized circuits.

### 4.1 "Deterministic errors" vs "errors produced by noise" vs "uncertainty errors"

First, notice that once we synthesize and use a circuit that computes a function $f'$ that approximates the original function $f$, we introduce "deterministic errors" into the computation. By this we mean that these errors cannot be undone anymore by using statistical methods or otherwise. This distinguishes them from what we refer to as "errors produced by noise" (errors that come from actual quantum hardware not behaving identical to the models of perfect quantum hardware) and "uncertainty errors" (errors due to the randomness inherent to quantum computations), which can be controlled by error-correction and probabilistic analysis. For algorithms where the oracle $f$ encodes the input-instance, using the approximate oracle $f'$ would effectively result in solving the problem for a different instance. Hence, if it is a problem where the solution is very sensitive to perturbations in the input-instance, our method would not be useful. For example, if we use the algorithm from Carette et al. [4] for detecting triangles in a graph specified by a graph-oracle, approximating the graph-oracle may give us a wrong solution because adding or removing a single edge from a graph can create or destroy a triangle.

On the other hand, if it is a problem where the solution is less sensitive to perturbations in the input-instance, it would be perfectly legitimate to approximate oracle-functions by other functions that can be computed more efficiently. For example, if $f$ encodes a data-set in a quantum machine learning algorithm, then $f$ is already an imperfect representation of the actual "underlying truth" (that is, $f$ already contains "deterministic errors" because of noise in the original data and working with a finite data-set), and hence the errors introduced by approximating $f$ would "mingle" with the errors that $f$ carries already. Also, when the oracle $f$ returns "continuous values" our approach would be useful since approximating $f$ would result in a "numerically close" function. This is for example the case in the work of Zhao et al. [37], where an oracle function is used to query a covariance matrix of a given data-set.

## 4.2 Optimizing exact implementations for highly used functions

As we have shown, our approach can also be used to find exact circuits, often beating the previous state-of-the-art circuits. Hence, for particular functions that are used among many algorithms we propose to use our algorithm to search new optimized circuits "offline"- that is, before compile time so that a compiler can store these circuits and use them when these functions are called. Examples of such functions include arithmetic functions like adders, multipliers and exponentiation as well as multi-controlled NOT-gates.

## 5 CONCLUSIONS AND FUTURE WORK

In this paper we studied the approximability of Boolean functions by reversible circuits with given (possibly insufficient) resources and architectural restrictions. We showed how EA can be used to find those reversible circuits. We obtained satisfactory results on all benchmark functions, indicating that a wide variety of functions can be well approximated by short reversible circuits. Experiments on a QC simulator show that under the influence of noise, approximate reversible circuits can have lower error-rates than exact circuits. In some cases, the error-rates of approximate circuits are more than 50% lower. This finding is of evident interest for QC. How approximate circuits can be integrated into more complex quantum algorithms is a non-trivial question that requires future work.

## REFERENCES
[1] 2020 (accessed May 19, 2020). *Device backend noise simulations*. https://qiskit.org/documentation/tutorials/simulators/2_device_noise_simulation.html
[2] Charles H. Bennett. 1989. Time/Space Trade-Offs for Reversible Computation. *SIAM J. Comput.* 18, 4 (1989), 766–776. https://doi.org/10.1137/0218053 arXiv:https://doi.org/10.1137/0218053
[3] Alex Bocharov, Yuri Gurevich, and Krysta M. Svore. 2013. Efficient decomposition of single-qubit gates into $V$ basis circuits. *Phys. Rev. A* 88 (Jul 2013), 012313. Issue 1. https://doi.org/10.1103/PhysRevA.88.012313
[4] Titouan Carette, Mathieu Laurière, and Frédéric Magniez. 2016. Extended Learning Graphs for Triangle Finding. https://doi.org/10.48550/ARXIV.1609.07786
[5] V. K. Chippa, S. T. Chakradhar, K. Roy, and A. Raghunathan. 2013. Analysis and characterization of inherent application resilience for approximate computing. In *2013 50th ACM/EDAC/IEEE Design Automation Conference (DAC)*. 1–9. https://doi.org/10.1145/2463209.2488873
[6] K. Datta, I. Sengupta, and H. Rahaman. 2012. Reversible circuit synthesis using evolutionary algorithm. In *2012 5th International Conference on Computers and Devices for Communication (CODEC)*. 1–4. https://doi.org/10.1109/CODEC.2012.6509351
[7] Christopher M. Dawson and Michael A. Nielsen. 2006. The Solovay-Kitaev Algorithm. *Quantum Info. Comput.* 6, 1 (Jan. 2006), 81–95.
[8] Leon Derczynski. 2016. Complementarity, F-score, and NLP Evaluation. In *Proceedings of the Tenth International Conference on Language Resources and Evaluation (LREC'16)*. European Language Resources Association (ELRA), Portorož, Slovenia, 261–266. https://www.aclweb.org/anthology/L16-1040
[9] Jay Gambetta and Sarah Sheldon. 2019 (accessed May 18, 2020). *Cramming More Power Into a Quantum Device*. https://www.ibm.com/blogs/research/2019/03/power-quantum-device/
[10] Adrian Gepp and Phil Stocks. 2007. A Review of Procedure to Evolve Quantum Procedures. (08 2007).
[11] N. Gleinig, F. A. Hubis, and T. Hoefler. 2019. Embedding Functions Into Reversible Circuits: A Probabilistic Approach to the Number of Lines. In *2019 56th ACM/IEEE Design Automation Conference (DAC)*. 1–6.
[12] Pawel Kerntopf. 2004. A New Heuristic Algorithm for Reversible Logic Synthesis. In *Proceedings of the 41st Annual Design Automation Conference (DAC '04)*. ACM, New York, NY, USA, 834–837. https://doi.org/10.1145/996566.996789
[13] M. H. A. Khan and M. Perkowski. 2004. Genetic algorithm based synthesis of multi-output ternary functions using quantum cascade of generalized ternary gates. In *Proceedings of the 2004 Congress on Evolutionary Computation (IEEE Cat. No.04TH8753)*, Vol. 2. 2194–2201 Vol.2. https://doi.org/10.1109/CEC.2004.1331169

[14] Sumeet Khatri, Ryan LaRose, Alexander Poremba, Lukasz Cincio, Andrew T. Sornborger, and Patrick J. Coles. 2019. Quantum-assisted quantum compiling. *Quantum* 3 (May 2019), 140. https://doi.org/10.22331/q-2019-05-13-140

[15] Vadym Kliuchnikov, Alex Bocharov, and Krysta M. Svore. 2014. Asymptotically Optimal Topological Quantum Compiling. *Phys. Rev. Lett.* 112 (Apr 2014), 140504. Issue 14. https://doi.org/10.1103/PhysRevLett.112.140504

[16] E. Knill. 2000. Approximation by Quantum Circuits. (08 2000).

[17] Dmitri Maslov and Gerhard W. Dueck. 2004. Reversible cascades with minimal garbage. *IEEE Trans. on CAD of Integrated Circuits and Systems* 23, 11 (2004), 1497–1509.

[18] Dmitri Maslov, Gerhard W Dueck, D Michael Miller, and Camille Negrevergne. 2008. Quantum circuit simplification and level compaction. *IEEE Transactions on Computer-Aided Design of Integrated Circuits and Systems* 27, 3 (2008), 436–444.

[19] Dmitri Maslov, Gerhard W. Dueck, and Nathan Scott. 2005 (accessed May 19, 2020). Reversible Logic Synthesis Benchmarks Page. http://webhome.cs.uvic.ca/ dmaslov.

[20] Dmitri Maslov, Christina Young, D. Miller, and G.W. Dueck. 2005. Quantum Circuit Simplification Using Templates. 1208–1213. https://doi.org/10.1109/DATE.2005.249

[21] Tzvetan S. Metodi and Frederic T. Chong. 2006. *Quantum Computing for Computer Architects (Synthesis Lectures on Computer Architecture).* Morgan and Claypool Publishers.

[22] D. M. Miller, D. Maslov, and G. W. Dueck. 2003. A transformation based algorithm for reversible logic synthesis. In *Proceedings 2003. Design Automation Conference (IEEE Cat. No.03CH37451).* 318–323. https://doi.org/10.1145/775832.775915

[23] Michael A. Nielsen and Isaac L. Chuang. 2010. *Quantum Computation and Quantum Information: 10th Anniversary Edition.* Cambridge University Press. https://doi.org/10.1017/CBO9780511976667

[24] John Preskill. 2018. Quantum Computing in the NISQ era and beyond. *Quantum* 2 (Aug 2018), 79. https://doi.org/10.22331/q-2018-08-06-79

[25] T. N. Sasamal, H. M. Gaur, A. K. Singh, and A. Mohan. 2020. Reversible Circuit Synthesis Using Evolutionary Algorithms. (2020), 115–128. https://doi.org/10.1007/978-981-13-8821-7_7

[26] L. Sekanina and Z. Vasicek. 2013. Approximate circuit design by means of evolvable hardware. In *2013 IEEE International Conference on Evolvable Systems (ICES).* 21–28. https://doi.org/10.1109/ICES.2013.6613278

[27] Peter Selinger. 2015. Efficient Clifford+T Approximation of Single-Qubit Operators. *Quantum Info. Comput.* 15, 1–2 (Jan. 2015), 159–180.

[28] Mathias Soeken, Robert Wille, Christoph Hilken, Nils Przigoda, and Rolf Drechsler. 2012. Synthesis of reversible circuits with minimal lines for large functions. In *Design Automation Conference (ASP-DAC), 2012 17th Asia and South Pacific.* IEEE, 85–92.

[29] Mathias Soeken, Robert Wille, Oliver Keszocze, D. Michael Miller, and Rolf Drechsler. 2015. Embedding of Large Boolean Functions for Reversible Logic. *J. Emerg. Technol. Comput. Syst.* 12, 4, Article 41 (Dec. 2015), 26 pages. https://doi.org/10.1145/2786982

[30] Zdenek Vasicek and Lukas Sekanina. 2015. Evolutionary Approximation of Complex Digital Circuits. In *Proceedings of the Companion Publication of the 2015 Annual Conference on Genetic and Evolutionary Computation (GECCO Companion '15).* ACM, New York, NY, USA, 1505–1506. https://doi.org/10.1145/2739482.2764657

[31] Milan Češka, Jiří Matyáš, Vojtech Mrazek, Lukas Sekanina, Zdenek Vasicek, and Tomas Vojnar. 2017. Approximating Complex Arithmetic Circuits with Formal Error Guarantees: 32-Bit Multipliers Accomplished. In *Proceedings of the 36th International Conference on Computer-Aided Design (ICCAD '17).* IEEE Press, 416–423.

[32] Swagath Venkataramani, Amit Sabne, Vivek Joy Kozhikkottu, Kaushik Roy, and Anand Raghunathan. 2012. SALSA: Systematic logic synthesis of approximate circuits. *DAC Design Automation Conference 2012* (2012), 796–801.

[33] R. Wille and R. Drechsler. 2009. BDD-based synthesis of reversible logic for large functions. In *2009 46th ACM/IEEE Design Automation Conference.* 270–275. https://doi.org/10.1145/1629911.1629984

[34] R. Wille, D. Große, L. Teuber, G. W. Dueck, and R. Drechsler. 2008. RevLib: An Online Resource for Reversible Functions and Reversible Circuits. In *Int'l Symp. on Multi-Valued Logic.* 220–225. RevLib is available at http://www.revlib.org.

[35] Robert Wille, Mathias Soeken, and Rolf Drechsler. 2010. Reducing the Number of Lines in Reversible Circuits. (2010), 647–652. https://doi.org/10.1145/1837274.1837439

[36] M. Zhang, S. Zhao, and X. Wang. 2009. Automatic synthesis of reversible logic circuit based on genetic algorithm. In *2009 IEEE International Conference on Intelligent Computing and Intelligent Systems,* Vol. 3. 542–546. https://doi.org/10.1109/ICICISYS.2009.5358132

[37] Zhikuan Zhao, Alejandro Pozas-Kerstjens, Patrick Rebentrost, and Peter Wittek. 2019. Bayesian deep learning on a quantum computer. *Quantum Machine Intelligence* 1, 1-2 (may 2019), 41–51. https://doi.org/10.1007/s42484-019-00004-7

[38] Bernard Zygelman. [n. d.]. *A First Introduction to Quantum Computing and Information.* Springer. https://doi.org/10.1007/978-3-319-91629-3